\documentclass[12pt]{amsart}
\usepackage{amsthm}
\usepackage{amstext}
\usepackage{amssymb}
\usepackage{mathrsfs}
\usepackage{mathabx}
\usepackage[english]{babel}
\usepackage{tikz-cd}
\usepackage{enumerate}
\usepackage{xfrac}
\usepackage{mathabx}
\usepackage{mathtools}
\usepackage{xcolor}
\usepackage[all]{xy}

\usepackage[T1]{fontenc}
\newcommand{\slim}{\mathop{s\rule[0.5ex]{1ex}{0.1ex}\mathrm{lim}}}

\theoremstyle{plain}
\newtheorem{theorem}{Theorem}[section]
\newtheorem{proposition}[theorem]{Proposition}

\newtheorem{lemma}[theorem]{Lemma}

\theoremstyle{definition}

\theoremstyle{remark}

\numberwithin{equation}{section}

\newcommand{\Id}{d}

\newcommand{\IC}{\mathbb{C}}
\newcommand{\IR}{\mathbb{R}}

\def\rm{\mathrm}
\def\bf{\mathbf}

\newcommand{\loc}{\mathrm{loc}}

\newcommand{\IHH}{\mathscr{H}}

\newcommand{\IN}{\mathbb{N}}

\renewcommand{\slim}{\mathop{\rm{st}\rule[0.5ex]{1ex}{0.1ex}\mathrm{lim}}}

\newcommand\newdot{{\kern.8pt\cdot\kern.8pt}}
\def\nbull{{\raise1.5pt\hbox{\bf .}}}

\title[Asymptotic Equivalence of Identification Operators]{Asymptotic Equivalence of Identification Operators in Geometric Scattering Theory}

\author{Batu Güneysu}

\begin{document}

\begin{abstract} Given two measures $\mu_1$ and $\mu_2$ on a measurable space $X$ such that $d\mu_2=\rho_{1,2} \, d\mu_1$ for some bounded measurable function $\rho_{1,2}:X\to (0,\infty)$, there exist two natural identification operators $J_{1,2},\tilde{J}_{1,2}:L^2(X,\mu_1)\to L^2(X,\mu_2)$, namely the unitary $J_{1,2}\psi:=\psi/\sqrt{\rho_{1,2}}$ and the trivial $\tilde{J}_{1,2}\psi:=\psi$. Given self-adjoint semibounded operators $H_j$ on $L^2(X,\mu_j)$, $j=1,2$, we prove a natural criterion in a topologic setting for the equality of the two-Hilbert-space wave operators $W_\pm(H_2,H_1;J_{1,2})$ and $W_\pm(H_2,H_1;\tilde{J}_{1,2})$, by showing that $J_{1,2}-\tilde{J}_{1,2}$ are asymptotically $H_1$-equivalent in the sense of Kato. It turns out that this criterion is automatically satisfied in typical situations on Riemannian manifolds and weighted infinite graphs in which one has the existence of completeness $W_\pm(H_2,H_1;\tilde{J}_{1,2})$ (and thus a-posteriori of $W_\pm(H_2,H_1;J_{1,2}))$.
\end{abstract}

\maketitle

\section{Introduction}

Geometric scattering theory has gained remarkable interest in the recent years \cite{muller,colin,keller, bei, baum,boldt, hempel, hempel2,thal, parra, ito} and usually takes place in a two-Hilbert-space setting: one starts with a measurable space $X$ and the two geometries in question on $X$ are encoded by pairs $(\mu_j,H_j)$, $j=1,2$, given by a measure $\mu_j$ on $X$ and a self-adjoint semibounded (from below) operator $H_j$ on $L^2(X,\mu_j)$. In typical applications the two measures are related by a positive Borel density function $\rho_{1,2}:X\to (0,\infty)$ according to $d\mu_2 = \rho_{1,2} \, d\mu_1$. For example, in the Riemannian setting, $\mu_j$ resp. $H_j$ is the volume measure resp. the Laplace-Beltrami operator induced by a (complete) Riemannian metric $g_j$ on a fixed noncompact manifold, noting that in this case $d\mu_2 /  d\mu_1$ is automatically a smooth positive function (cf. Section \ref{m}). By letting $g_2$ run, one can obtain stability results for the absolutely continuous spectrum under the Ricci flow \cite{thal} in this way. Likewise one can consider two infinite weighted graphs $(X,b_j,\mu_j)$ with $H_j$ the induced weighted Laplacian in $\ell^2(X,\mu_j)$, where again $d\mu_2 /  d\mu_1$ becomes a positive function (cf. Section \ref{g}).\vspace{1mm}

Coming back to  the general setting of measures spaces, in order to formulate that the difference of $H_2$ and $H_1$ is a scattering obstacle of $H_1$, one needs a bounded identification operator $J:L^2(X,\mu_1)\to L^2(X,\mu_2)$, which allows to define the two-Hilbert-space wave operators $W_\pm(H_2,H_1;J)$ (cf. Section \ref{eins}). One natural choice (which does not require any additional assumptions) is the unitary identification operator $J_{1,2}:L^2(X,\mu_1)\to  L^2(X,\mu_2)$ given by $J_{1,2}\psi:= \psi/\sqrt{\rho_{1,2}}$. This identification is used for example in \cite{parra} for scattering the geometry of some infinite weighted graphs. However, some recently established analytic techniques (called \emph{Hempel-Post-Weder decomposition formula} in \cite{thal,keller}, based on \cite{hempel}) that are used to prove the existence and completeness of the wave operators require instead that $\rho_{1,2}$ is bounded and to work with the bounded identification operator
$$
\tilde{J}_{1,2}:L^2(X,\mu_1)\longrightarrow  L^2(X,\mu_2),\quad \tilde{J}_{1,2}\psi:= \psi
$$
instead of $J_{1,2}$. This raises the following natural question: under which assumptions does one have the equality 
\begin{align}\label{aaas}
W_\pm(H_2,H_1;J_{1,2})=W_\pm(H_2,H_1;\tilde{J}_{1,2})
\end{align}
of the induced wave operators? At an abstract level, this question is answered by Kato in \cite{kato}: namely, the equality (\ref{aaas}) is equivalent to $J_{1,2}$ and $\tilde{J}_{1,2}$ being \emph{asymptotically $H_1$-equivalent} (cf. Section \ref{eins}). \vspace{1mm}

In order to formulate our main result, assume now that $X$ is a locally compact and seperable metrizable space and that the $\mu_j$ are fully supported Radon measures on $X$. We assume also that the heat semigroup of $H_1$ has the smoothing property
$$
\exp(-sH_1):L^2(X,\mu_1)\longrightarrow C(X)\quad\text{for all $0<s<1$},
$$
which is a natural assumption in geometric scattering theory, for together with Grothendieck's factorization trick it implies that all $H_1$-absolutely continuous states are scattering states (cf. Proposition \ref{rage}). Then with the Borel function 
$$
\phi_{1}(s,\cdot):X\longrightarrow [0,\infty),\quad \phi_{1}(s,x):=\sup_{\psi}\frac{|\exp(-sH_1)\psi(x)|^2}{ \int |\psi(y)|^2 d\mu_1(y)}
$$
our abstract main result, Theorem \ref{main}, states that $J_{1,2}$ and $\tilde{J}_{1,2}$ are asymptotically $H_1$-equivalent, provided
\begin{align}\label{assss}
\int (1-1/\rho_{1,2}(x))^2 \phi_{1}(s,x) \, d\mu_1(x)<\infty.
\end{align}
It came as a surprise to us that the assumptions of the criteria for existence and completeness of $W_\pm(H_2,H_1;\tilde{J}_{1,2})$ on noncompact Riemannian manifolds \cite{thal} and infinite weighted graphs \cite{keller} (which both contain the current state-of-the-art) automatically imply (\ref{assss}) and thus (\ref{aaas}). This is clarified in Section \ref{m} and Section \ref{g}, respectively.\vspace{1mm}

Finally, we remark that the techniques developed here can be generalized to geometric scattering theory on vector bundles, as no genuinely scalar assumptions are made on the $H_j$'s (such as e.g. that $\exp(-t H_j)$ positivity preserving). The vector bundle setting arises naturally \cite{bei, baum} in the scattering theory of the Hodge-Laplacian and \cite{boldt} the scattering theory of the Dirac operator and its square, noting that in the Dirac case even the vector bundles themselves change with the geometry. We have not addressed this issue here as the formulation of these results would be somewhat cumbersome, without producing new analytic insights.\vspace{3mm}

\textbf{Acknowledgements:} The author would like to thank Sebastian Boldt, Ognjen Milatovic, and Peter Stollmann (from whom he learned about Grothendieck's factorization trick).

\section{Two-Hilbert-Space Scattering Theory}\label{eins}

Given a self-adjoint operator $H$ in the complex Hilbert space $\IHH$, let $E_H$ denote the projection-valued spectral measure of $H$ and let $\pi^{\mathrm{ac}}_H$ denote the projection on the closed subspace of $\IHH$ given by all $\psi\in \IHH$ such that the Borel probability measure $\left\|E_H(\cdot)\psi\right\|^2$ on $\IR$ is absolutely continuous with respect to the Lebesgue measure. For future reference we record the following well-known fact, which is ultimately a simple consequence of the Riemannian-Lebesgue Lemma applied to the spectral calculus:

\begin{lemma}\label{simp} Let $D:\IHH\to \tilde{\IHH} $ be a compact operator from $\IHH$ to some complex Hilbert space $\tilde{\IHH}$ such that $D E_H(-l,l)$ is compact for every interval $(-l,l)$, $l\in\IN$. Then for every $\psi\in \mathrm{Ran}(\pi^{\mathrm{ac}}_H)$ one has 
\begin{align*}
\| D\exp(-itH)\psi \|\to 0\text{ as $t\to\pm\infty$}.
\end{align*}
\end{lemma}

\begin{proof} Lemma 2 on page 24 in \cite{rs2} states that $\big\| \tilde{D}\exp(-itH_1)\psi\big\|\to 0$ as $t\to\pm\infty$ for every compact operator $\tilde{D}$. In view of this, we estimate as follows:
\begin{align*}
	&\big\| D\exp(-itH)\psi\big\|\\
	&\leq \big\|D\exp(-itH)\psi-DE_H(-l,l)\exp(-itH)\psi\big\|+\big\|DE_H(-l,l)\exp(-itH)\psi\big\|\\
	&=\big\|D\exp(-itH)\psi-D\exp(-itH)E_H(-l,l)\psi\big\| +\big\|DE_1(-l,l)\exp(-itH)\psi\big\|\\
	&=\big\|D\exp(-itH)\big(\psi-E_H(-l,l)\psi\big)\big\|+\big\|DE_H(-l,l)\exp(-itH)\psi\big\|\\
	&\leq \left\|D\right\|\big\|\psi-E_H(-l,l)\psi\big\|+\big\|DE_H(-l,l)\exp(-itH)\pi\big\|.
\end{align*}
The second summand one goes to zero as $t\to\pm\infty$ by the above fact, and the first summand goes to zero as $l\to\infty$ by spectral calculus.
\end{proof}

Assume we are given a self-adjoint operator $H_j$ in the complex Hilbert space $\IHH_j$ for $j=1,2$, with $E_j:=E_{H_j}$ and $\pi^{\mathrm{ac}}_j:=\pi^{\mathrm{ac}}_{H_j}$. Given a bounded operator $J:\IHH_1\to\IHH_2$ one says that the \emph{wave operators $W_\pm(H_2,H_1;J)$ exist}, if the strong limits 
$$
W_\pm(H_2,H_1;J):=\slim_{t\to \pm\infty} \exp(itH_2)J\exp(-itH_1)\pi_1^{\mathrm{ac}}\\
$$
exist, in which case one calls $W_\pm(H_2,H_1;J)$ \emph{complete}, if
$$
\mathrm{Ker}\big(W_\pm(H_2,H_1;J))^\perp=\mathrm{Ran}(\pi_1^{\mathrm{ac}}),\quad \overline{\mathrm{Ran}\big(W_\pm(H_2,H_1;J))}=\mathrm{Ran}(\pi_2^{\mathrm{ac}}).
$$
Recalling that the absolutely continuous part $H^\mathrm{ac}_j$ of $H_j$ is defined as the self-adjoint operator obtained by restricting $H_j$ to the Hilbert space $\mathrm{Ran}(\pi_j^{\mathrm{ac}})$, the completeness of $W_\pm(H_2,H_1;J)$ implies \cite{kato} that $H^\mathrm{ac}_1$ is unitarily equivalent to $H^\mathrm{ac}_2$, which implies that the absolutely continuous spectra coincide:
$$
\mathrm{spec}_{\mathrm{ac}}(H_1):=\mathrm{spec}(H_1^{\mathrm{ac}})=\mathrm{spec}(H_2^{\mathrm{ac}})=:\mathrm{spec}_{\mathrm{ac}}(H_2).
$$
Given two bounded operators $\tilde{J},J:\IHH_1\to\IHH_2$, following Kato \cite{kato}, one says that $\tilde{J}$ and $J$ are \emph{asymptotically $H_1$-equivalent}, if 
$$
\slim_{t\to \pm\infty} (\tilde{J}-J)\exp(-itH_1)\pi^{\mathrm{ac}}_1=0.
$$
It follows \cite{kato} that if $\tilde{J}$ and $J$ are asymptotically $H_1$-equivalent, then one has $W_\pm(H_2,H_1;\tilde{J})=W_\pm(H_2,H_1;J)$ in the sense that the existence of one side implies the existence of the other and equality.

\section{Two-Hilbert-Space Scattering on Locally Compact Spaces}\label{zwei}

Assume now $X$ is a locally compact seperable metrizable space and $\mu$ is a Radon measure on $X$ with full support. If $H$ is a self-adjoint operator in $L^2(X,\mu)$ which is semibounded (from below) and if $s>0$ is such that
\begin{align}\label{kernel}
	\text{$\mathrm{Ran}\big(\exp(-sH)\big)\subset  C(X)$},
\end{align}
then the induced operator from $L^2(X,\mu)$ to $C(X)$ is automatically bounded by the closed graph theorem, when $C(X)$ is turned into a Fréchet space with respect to uniform convergence on compacts (noting that $X$ is compactly exhaustable). In particular, for all $s>0$, $K\subset X$ compact there exists a constant $c=c(H,s,K)>0$ with 
$$
\sup_K|\exp(-sH)\psi|^2\leq c\int |\psi(y)|^2 d\mu(y) \quad \text{for all $\psi\in L^2(X,\mu)$}.
$$

It then follows from the seperability of $L^2(X,\mu)$ that the function
\begin{align}\label{conti}	
\phi_{H}(s,\cdot): X\longrightarrow [0,\infty),\quad \phi_{H}(s,x):=\sup_{\psi}\frac{|\exp(-sH)\psi(x)|^2}{ \int |\psi(y)|^2 d\mu(y)}
\end{align}
is a well-defined Borel function. The importance of a smoothing assumption such as (\ref{kernel}) in the context of scattering theory is reflected by the fact that every $H$-absolutely continuous initial state $\psi$ is a scattering state: 

\begin{proposition}\label{rage} Assume $H$ is a self-adjoint semibounded operator in $L^2(X,\mu)$ which satisfies (\ref{kernel}) for some $s>0$. Then for every $\psi\in \mathrm{Ran}(\pi^{\mathrm{ac}}_H)$ and every compact $K\subset X$ one has
	$$
	\lim_{t\to\pm\infty}\int_{K} \left|\exp(-it H)\psi(x) \right|^2 \, d\mu(x) =0.
	$$
\end{proposition}

\begin{proof} This is a simple consequence of Grothendieck's factorization trick (cf. \cite{peter} for a very nice discussion of this result), which states that every operator $A$ in $L^2(X,\mu)$ which factors boundedly according to the diagram
	$$
	\begin{tikzcd}[column sep=small]
		L^2(X,\mu) \arrow{rr}{A} \arrow[swap]{dr}{A_1}& & L^2(X,\mu) \\
		& L^\infty(X,\mu) \arrow{ur}[swap]{A_2} & 
	\end{tikzcd}    
		$$
	is Hilbert-Schmidt. Let $M_K$ denote the multiplication operator in $L^2(X,\mu)$ induced by $1_K$. The operator 
$$	
M_{K}\exp(-s H)=M_{K}M_{K}\exp(-s H) \>\>\>\text{factors boundedly according to}\>\>\>   M_{K}\exp(-s H)=A_2A_1, 
$$
where 
	\begin{align*}
&	A_1:L^2(X,\mu)\longrightarrow L^\infty(X,\mu),\quad A_1\psi:= 1_{K} \exp(-s H)\psi,\\
&	A_2:L^\infty(X,\mu)\longrightarrow L^2(X,\mu),\quad A_2\psi:= 1_{K}  \psi.
\end{align*}	
	Thus by Grothendieck's factorization trick the operator $M_{K}\exp(-s H)$ is Hilbert-Schmidt and thus compact. It follows that 
	$$
	M_{K} E_H(-l,l)=M_{K} \exp(-s H) \exp(s H) E_H(-l,l)
	$$ 
	is compact for every bounded interval $(-l,l)$, which proves the claim by Lemma \ref{simp}.
\end{proof}

If a quantum particle has the energy operator $H$ and an initial state $\psi\in L^2(X,\mu)$ with $\int|\psi(x)|^2 d\mu(x)=1$, then $\psi(t,x):=\exp(-it H)\psi(x)$ is its state at time $t$, and the probability of finding the particle in a Borel set $Y\subset X$ at the time $t$ is given by $\int_{Y} |\psi(t,x) |^2 d\mu(x)$. Thus Proposition \ref{rage} shows that if the initial state is an $H$-absolutely continuous one, then the particle will eventually leave every compact set.\vspace{2mm}


We fix now two fully supported Radon measures $\mu_j$ on $X$, $j=1,2$, as well as two self-adjoint semibounded operators $H_j$ in $L^2(X,\mu_j)$. We assume
$$
\Id\mu_2 = \rho_{1,2} \, \Id\mu_1\quad\text{for some Borel function $\rho_{1,2}:X\to (0,\infty)$}.
$$
Then there is a unitary map
\begin{align*}
J_{1,2}:L^2(X,\mu_1)\longrightarrow L^2(X,\mu_2),\quad J_{1,2}\psi:= \psi/\sqrt{\rho_{1,2}}.
\end{align*}
If $\rho_{1,2}$ is bounded, we also get the bounded map
$$
\tilde{J}_{1,2}:L^2(X,\mu_1)\longrightarrow L^2(X,\mu_2),\quad \tilde{J}_{1,2}\psi:= \psi.
$$
We set 
$$
\phi_{j}(s,x):=\phi_{H_j}(s,x),
$$
whenever this function is well-defined. Here comes our main result:


\begin{theorem}\label{main} Assume $0< \inf\rho_{1,2}\leq\sup  \rho_{1,2}<\infty$ as well as (\ref{kernel}) and 
\begin{align}\label{asp}
\int ( 1-  1/\rho_{1,2} (x)   )^2 \phi_{1}(s,x)\, d\mu_1(x)<\infty\quad\text{for all $0<s<1$}.
\end{align}
Then $\tilde{J}_{1,2}$ and $J_{1,2}$ are asymptotically $H_1$-equivalent, in particular, 
$$
W_\pm(H_2,H_1;J_{1,2})=W_\pm(H_2,H_1;\tilde{J}_{1,2}).
$$	
\end{theorem}	

\begin{proof} Set $\rho:=\rho_{1,2}$, $J:=J_{1,2}$, $\tilde{J}:=\tilde{J}_{1,2}$. By the semigroup property (\ref{kernel}) holds for all $s>0$ and then the proof of Proposition 3.3 from \cite{lenz} entails the existence of a uniquely determined jointly Borel measurable map
$$
(0,\infty)\times X\times X \ni (r,x,y)\longmapsto \exp(-rH_1)(x,y)\in\IC,
$$
such that for all $r>0$, $x\in X$ one has $\exp(-rH_1)(x,\cdot)\in L^2(X,\mu_1)$ with
$$
\exp(-rH_1)\psi(x)=\int \exp(-rH_1)(x,y)\psi(y)\, \Id \mu_1( y)\quad\text{for all $\psi\in L^2(X,\mu_1)$.}
$$
Let us first show that $(\tilde{J}-J)\exp(-sH_1)$ is compact: define a bounded bijective operator $A$ with bounded inverse via
	\begin{align*}
		&A:L^2(X,\mu_2)\longrightarrow L^2(X,\mu_2),\quad A\psi:= (1+\rho^{-1/2})\psi,\quad\text{so that}\\
		&A^{-1}:L^2(X,\mu_2)\longrightarrow L^2(X,\mu_2),\quad A^{-1}\psi= (1+\rho^{-1/2})^{-1}\psi,	
	\end{align*}
noting that the boundedness of $A$ follows from $\inf\rho>0$. The assumption (\ref{asp}) implies that the operator $A(\tilde{J}-J)\exp(-sH_1)$ is a Hilbert-Schmidt integral operator: indeed, its integral kernel is given by
 $$
	[A(\tilde{J}-J)\exp(-sH_1)](x,y)=\big(1-\rho(x)^{-1}\big)\exp(-sH_1)(x,y),
	$$	
and by the very definition of $\phi_{1}(s,x)=\phi_{H_1}(s,x)$ and Riesz-Fischer duality,
$$
\int|\exp(-sH_1)(x,y)|^2 \, d\mu_1(y)\leq \phi_{1}(s,x),
$$
and so
\begin{align*}
&\int\int\big|[A(\tilde{J}-J)\exp(-sH_1)](x,y)\big|^2\, d\mu_1(y)\, d\mu_1(x)\\
&=\int\big(1-\rho(x)^{-1}\big)^2\int|\exp(-sH_1)(x,y)|^2\, d\mu_1(y)\, d\mu_1(x)\\
&\leq \int \big( 1-  \rho (x)^{-1}  \big)^2 \phi_{1}(s,x)\, \Id\mu_1(x)<\infty.
 \end{align*}
In particular, $A(\tilde{J}-J)\exp(-sH_1)$ is compact for all $0<s<1$, and thus 
	$$
    (\tilde{J}-J)\exp(-sH_1)=A^{-1}A (\tilde{J}-J)\exp(-sH_1)	
	$$
is compact, too. It follows that 
$$
(\tilde{J}-J)E_1(-l,l)=   (\tilde{J}-J)\exp(-sH_1)\exp(sH_1)E_1(-l,l)
$$ 
is compact for every bounded interval $(-l,l)$, $l\in\IN$, so that the claim follows from Lemma \ref{simp}.

\end{proof}

\section{Application to Noncompact Riemannian Manifolds}\label{m}

Let $X$ be a connected noncompact manifold (smooth, without boundary) of dimension $m$ and let $g$ be a (smooth) Riemannian metric on $X$, that is, $g$ a smooth $(0,2)$ tensor field on $X$ with $g(x)$ a scalar product on each tangent space $T_x X$. Let $\mu_g$ denote the induced Riemannian volume measure. This measure is uniquely determined by its local values on charts, where with the symmetric strictly positive definite matrix  $\mathbf{g}_{jk}:=g(\partial_j,\partial_k)$ one has
\begin{align}\label{local}
\Id\mu_g(x)= \sqrt{\det(\mathbf{g})}\,\Id x.
\end{align}
The metric $g$ allows to calculate \cite{grig} the length $L_g(\gamma)$ of a smooth curve $\gamma:[0,1]\to X$ according to
$$
L_g(\gamma):=\int^1_0 \sqrt{g(\dot{\gamma}(s),\dot{\gamma}(s))}\,ds,
$$
which incudes a distance function on $X$ via
$$
d_g(x,y):=\{ L_g(\gamma):\text{$\gamma$ is smooth curve from $x$ to $y$}\}.
$$

This distance induces the original topology on $X$, its induced open balls are denoted with $B_g(x,r)$, and finally
$$
\mu_g(x,r):=\mu_g(B_g(x,r))
$$
stands for the volume function. The symbol $\Delta_{g}$ denotes the negative-definite Laplace-Beltrami operator, which is the second order elliptic operator given locally by
$$
\Delta_g= \sum^m_{j,k=1}\frac{1}{\sqrt{\det(\mathbf{g})}}\partial_j \circ \sqrt{\det(\mathbf{g})}\mathbf{g}_{jk}\partial_k,
$$ 
with $\mathbf{g}_{jk}$ the components of $\mathbf{g}^{-1}$. The symbol $H_g\geq 0$ stands for the Friedrichs realization of the symmetric nonnegative operator $-\Delta_g$ in $L^2(X,\mu_g)$, defined a priori on smooth compactly supported functions. In other words, the nonnegative sesqulinear form in $L^2(X,\mu_g)$ given by
$$
C^\infty_c(X)\times C^\infty_c(X) \ni (\psi_1,\psi_2)\longmapsto -\int \overline{\Delta_g\psi_1(x)} \, \psi_2(x) \, d\mu_g(x)\in \IC
$$ 
is closable and $H_g$ is induced by the closure of this form. If $g$ (that is, $d_g$) is complete, then $-\Delta_g$ is essentially self-adjoint \cite{strichartz}. \\
The assumption (\ref{kernel}) is clearly always satisfied for all $s>0$, as 
$$
\mathrm{Ran}\big(\exp(-s H_g)\big)\subset \bigcap_{n\in \IN} \mathrm{Dom}(H_g^n)\subset \bigcap_{n\in \IN} W^{2n,2}_\loc(X)\subset C^\infty(X),
$$
where the first inclusion follows from spectral calculus, the second one from local elliptic regularity and the last one from Sobolev's embedding theorem. In fact, by local parabolic regularity, $s\mapsto \exp(-s H_g)$ has a nonnegative integral kernel which is jointly smooth in $(s,x,y)$ and Markovian (cf. Theorem 7.13 in \cite{grig}):
\begin{align}\label{markoff}
\int \exp(-s H_g)(x,y) d\mu_g(y)\leq 1\quad\text{for all $s>0$.}
\end{align}
Finally, the Ricci curvature $\mathrm{Ric}^g$ is the smooth $(0,2)$ tensor field on $X$ given locally in terms of the Christoffel symbols
$$
\Gamma^{g;k}_{pj}:=\frac{1}{2}\sum^m_{l=1}(\mathbf{g}^{kl}\partial_p \mathbf{g}_{jl}+\mathbf{g}^{kl}\partial_j\mathbf{g}_{pl}-\mathbf{g}^{kl}\partial_l \mathbf{g}_{pj})
$$
by 
$$
\mathrm{Ric}^g_{jk}=\sum^m_{l=1}(\partial_{l} \Gamma^{g;l}_{jk}  - \partial_{j}\Gamma^{g;l}_{kl})+\sum^m_{p,l=1}(\Gamma^{g;l}_{lp} \Gamma^{p}_{jk}-\Gamma^{g;l}_{jp} \Gamma^{g;p}_{lk}).
$$
One says that the Ricci curvature of $g$ is bounded from below, if
$$
\inf_{x\in X}\inf_{ v\in T_x X\setminus \{0\}}\frac{\mathrm{Ric}^g(x)(v,v)}{g(x)(v,v)}>-\infty.
$$

Fix two Riemannian metrics $g_1$ and $g_2$ on $X$ and set $\mu_j:=\mu_{g_j}$, $H_j:=H_{g_j}$. We can always write 
$$
\Id\mu_2= \rho_{1,2} \,\Id\mu_1,
$$ 
where the smooth density function $\rho_{1,2}:X\to (0,\infty)$ is given locally by 
$$
\rho_{1,2}:=\sqrt{\det(\mathbf{g}_{2} )   }/\sqrt{\det(\mathbf{g}_{1})},
$$ 
noting that although each $\sqrt{\det(\mathbf{g}_{j})}$ is only locally defined, their quotient is globally defined. We get the unitary operator
$$
J_{1,2}:L^2(X,\mu_1)\longrightarrow L^2(X,\mu_2),\quad J_{1,2}\psi:= \psi/\sqrt{\rho_{1,2}}.
$$
Assume now that $g_1$ is quasi-isometric to $g_2$, meaning that there exists a constant $a>0$ with
$$
(1/a)g_1(x)(v,v)\leq g_2(x)(v,v)\leq ag_2(x)(v,v)\quad\text{for all $x\in X$, $v\in T_x X$.} 
$$ 
Then we have $0<\inf \rho_{1,2}\leq \sup \rho_{1,2}<\infty$, so there is another bounded operator
$$
\tilde{J}_{1,2}:L^2(X,\mu_1)\longrightarrow L^2(X,\mu_2),\quad \tilde{J}_{1,2}\psi:= \psi.
$$
Let $\mathscr{A}_{1,2}$ be the smooth $(1,1)$ tensor field on $X$ which is given locally by $\mathbf{g}_{1}\mathbf{g}^{-1}_2$. In particular, $\mathscr{A}_{1,2}(x):T_xX\to T_x X$ has strictly positive eigenvalues for all $x\in X$ and we can define the Borel function
$$
\delta_{1,2}:X\longrightarrow [0,\infty), \quad \delta_{1,2}(x):=2\sinh\Big((m/4)\max_{\lambda\in \mathrm{spec}(\mathscr{A}_{1,2}(x))}|\log(\lambda)|\Big).
$$
As an application of Theorem \ref{main} we obtain:

\begin{theorem}\label{riem} Let $g_1$ and $g_2$ be quasi-isometric and complete, both with Ricci curvature bounded from below. Assume further that for some (and then by quasi-isometry: both) $j=1,2$ and all $0<s<1$, 
\begin{align}\label{anm}
\int \delta_{1,2}(x)\mu_j(x,s)^{-1} \, \Id \mu_j( x)<\infty.
\end{align}
Then $W_\pm(H_2,H_1;J_{1,2})=W_\pm(H_2,H_1;\tilde{J}_{1,2})$, including existence and completeness. 	
\end{theorem}

\begin{proof} Set $\rho:=\rho_{1,2}$, $\tilde{J}:=\tilde{J}_{1,2}$, $J:=J_{1,2}$. It has been shown in \cite{thal} that under the given assumptions $W_\pm(H_2,H_1;\tilde{J})$ exist and are complete. It remains to check (\ref{asp}): to this end, we remark that by Li-Yau's heat kernel upper bound \cite{li},
	$$
	\sup_{y\in X}\exp(-sH_1)(x,y)\leq  C\mu_1(x,s)^{-1} \quad\text{for all $x\in X$, $0<s<1$,}
	$$
where the constant $C$ depends only on the dimension $m$ and the lower Ricci curvature bound of $g_1$. Thus for all $\psi\in L^2(X,\mu_1)$, using Cauchy-Schwarz and (\ref{markoff}),
\begin{align*}
&|\exp(-sH_1)\psi(x)|^2\\
&\leq \left(\int \exp(-sH_1)(x,y)|\psi(y)|\,d\mu_1(y)\right)^2\\
&\leq \int \exp(-sH_1)(x,y)^2d\mu_1(y)  \int |\psi(y)|^2\,d\mu_1(y)\\
&\leq C \mu_1(x,s)^{-1} \int |\psi(y)|^2\,d\mu_1(y).
\end{align*}
Recalling the definition (\ref{conti}) of $\phi_{1}(s,x)=\phi_{H_1}(s,x)$, it follows that $\phi_{1}(s,x)\leq C \mu_1(x,s)^{-1}$ and so
\begin{align*}
&\int( 1-  \rho(x)^{-1}   )^2\phi_{1}(s,x)\,\Id \mu_1(x )\\
&\leq C (1+\left\|1/\rho\right\|_\infty)\int\big| 1-  \rho(x)^{-1}   \big|\mu_1(x,s)^{-1}\,\Id \mu_1(x )\\
&= C (1+\left\|1/\rho\right\|_\infty)\int|\rho(x)^{-1/2}|\big| \rho(x)^{1/2}-  \rho(x)^{-1/2}   \big|\mu_1(x,s)^{-1}\,\Id \mu_1(x )\\
&\leq  C (1+\left\|1/\rho\right\|_\infty)\left\|1/\sqrt{\rho}\right\|_\infty\int\big| \rho(x)^{1/2}-  \rho(x)^{-1/2}   \big|\mu_1(x,s)^{-1}\,\Id \mu_1(x ).
\end{align*}
Lemma 3.3 in \cite{hempel} states that
$$
|\rho(x)^{1/2}-  \rho(x)^{-1/2}|\leq \delta(x),
$$
which with $\mathscr{A}:=\mathscr{A}_{1,2}$ is in fact a simple consequence of 
$$
\rho(x)^{1/2}-\rho(x)^{-1/2}=2\sinh\big((1/2)\log(\rho(x))\big),\quad \rho(x)=\det(\mathscr{A}(x))^{-1/2}.
$$
We arrive using (\ref{anm}) at
\begin{align*}
\int\big( 1-  \rho(x)^{-1}   \big)^2\phi_{1}(s,x)\,\Id \mu_1(x )<\infty,
\end{align*}
completing the proof.
\end{proof}

\section{Application to Weighted Infinite Graphs}\label{g}

Let the triple $(X,b,\mu)$ be a \emph{weighted infinite graph}, that is, $X$ is an infinite countable set (with the discrete topology) with $\mu:X\to (0,\infty)$ a function and $b:X\times X\to [0,\infty)$ a symmetric function which is zero on the diagonal and satisfies
$$
\sum_{y\in X} b(x,y)<\infty\quad\text{ for all $x\in X$}.
$$
Here $X$ is considered the vertices of a graph, $\mu$ as a vertex weight function, and $b$ as an edge weight function, where $\{b>0\}$ are the edges of a graph. The function $\mu$ induces a fully supported measure on $X$ which with the usual abuse of notation is given by $\mu(A)=\sum_{x\in A}\mu(x)$. The nonnegative sesquilinear form in $\ell^2(X,\mu)$ defined on finitely supported functions by
$$
C_c(X)\times C_c(X)\ni(\psi_1,\psi_2) \longmapsto \sum_{x\in X, y\in X}b(x,y)   \overline{\big(\psi_1(x)-\psi_2(x)\big)}\big(\psi_1(y)-\psi_2(y)\big)   \in\IC
$$
is closable by Fatou's lemma. Its closure thus induces a nonnegative (in general unbounded) self-adjoint operator $H_{b,\mu}$ in $\ell^2(X,\mu)$ which plays the role of the Laplacian in this context. In fact, if $(X,b)$ is locally finite, meaning that 
$$
\# \{y\in X: b(x,y)>0\}<\infty\quad\text{for all $x\in X$},
$$
then one has
$$
H_{b,\mu}\psi(x)=\frac{1}{\mu(x)}\sum_{\{y:b(x,y)>0\}} b(x,y)(\psi(x)-\psi(y))\quad\text{for all $\psi\in C_c(X)$.}
$$

By discreteness every bounded operator satisfies (\ref{kernel}). In fact, $\exp(-sH_{b,\mu})$ has a nonnegative integral kernel which satisfies
\begin{align}\label{schlau}
\sum_{x\in X}\exp(-sH_{b,\mu})(x,y) \mu(y)\leq 1\quad\text{ for all $ x\in X$, $s>0$}.
\end{align}
We refer the reader to \cite{book} for the foundations of such weighted graph Laplacians.\vspace{1mm}

Fix now two such weighted infinite graphs $(X,b_j,\mu_{j})$, $j=1,2$, and set $H_{j}:=H_{b_j,\mu_j}$. Clearly
$$
d\mu_2 = \rho_{1,2} \, d\mu_1,\quad\text{ with $\rho_{1,2}:=\mu_1(x)/\mu_2(x)$,}
$$
and there is the unitary operator
$$
J_{1,2}:\ell^2(X,\mu_1)\longrightarrow \ell^2(X,\mu_2),\quad J_{1,2}\psi:=\psi/\sqrt{\rho_{1,2}}.
$$
If we assume $\mu_1\sim \mu_2$ in the sense of $(1/a) \mu_1\leq \mu_2\leq a\mu_1$ for some constant $a>0$, then we have $0<\inf\rho_{1,2}\leq \sup\rho_{1,2}<\infty$ and we also get the bounded operator
$$
\tilde{J}_{1,2}: \ell^2(X,\mu_1)\longrightarrow \ell^2(X,\mu_2),\quad \tilde{J}_{1,2}\psi:=\psi.
$$
Define
$$
\tilde{\rho}_{1,2}:X\times X\longrightarrow [0,\infty),\quad \tilde{\rho}_{1,2}(x,y):= \begin{cases}&b_1(x,y)/b_2(x,y),\text{ if $b_2(x,y)\ne 0$}\\
	&1, \text{else}.
\end{cases}
$$
In analogy to the $\mu_j$'s, the notation $b_1\sim b_2$ means that $(1/a) b_1\leq b_2\leq ab_1$ for some constant $a>0$.

\begin{theorem}\label{main2} Assume 
	\begin{align}
\label{erst}		
&\mu_1\sim \mu_2,        \quad \sum_{x\in X}\big|\rho_{1,2}(x)^{\frac{1}{2}}-\rho_{1,2}(x)^{-\frac{1}{2}}\big| <\infty, \\
&b_1\sim b_2,\quad 		\sum_{x\in X}\sum_{y\in X}\big|\tilde{\rho}_{1,2}(x,y)^{\frac{1}{2}}-\tilde{\rho}_{1,2}(x,y)^{-\frac{1}{2}}\big|\big(\mu_j(x)^{-1}+\mu_j(y)^{-1}\big)b_j(x,y)  <\infty,\quad j=1,2.
	\end{align}
	Then one has $W_{\pm}(H_{2},H_1, J_{1,2})=W_{\pm}(H_{2},H_1, \tilde{J}_{1,2})$, including existence and completeness.
\end{theorem}

\begin{proof} Set $\rho:=\rho_{1,2}$, $\tilde{J}:=\tilde{J}_{1,2}$, $J:=J_{1,2}$. It has been shown in \cite{keller} that under the given assumptions the $W_{\pm}(H_{2},H_1, \tilde{J})$ exist and are complete. Let us show that (\ref{erst}) implies (\ref{asp}): it follows from (\ref{schlau}) that
$$
\sup_{y\in X}\exp(-sH_1)(x,y)\leq  \mu_1(x)^{-1} \quad\text{for all $x\in X$, $s> 0$,}
$$
which using Cauchy-Schwarz and (\ref{schlau}) as in the proof of Theorem \ref{riem} implies $\phi_{1}(s,x)\leq \mu_1(x)^{-1}$, and
$$
\sum_{x\in X}( 1-\rho(x)^{-1}   )^2\phi_{1}(s,x)\mu_1(x)\leq C_1	\sum_{x\in X}| 1-\rho(x)^{-1}   |\leq  C_2\sum_{x\in X}|\rho(x)^{\frac{1}{2}}-\rho(x)^{-\frac{1}{2}}\big| <\infty,
$$
completing the proof.

\end{proof}

\end{document}